\documentclass[aps,prl,preprint,superscriptaddress,showpacs,byrevtex]{revtex4}
\usepackage{graphicx} 
\usepackage{dcolumn}  

\graphicspath{{ps}}

\begin{document}
\newcommand{\Bo}{B^{0}}
\newcommand{\Bp}{B^{+}}
\newcommand{\Bm}{B^{-}}
\newcommand{\Dp}{D^{+}}
\newcommand{\Dm}{D^{-}}
\newcommand{\Do}{D^{0}}
\newcommand{\Dob}{\overline{D}^{0}}
\newcommand{\Dst}{D^{*}}
\newcommand{\Dsto}{D^{*0}}
\newcommand{\Dstb}{\overline{D}^{*}}
\newcommand{\Dstp}{D^{*+}}
\newcommand{\Dstm}{D^{*-}}
\newcommand{\DDst}{DD^{*}}
\newcommand{\Kp}{K^{+}}
\newcommand{\Km}{K^{-}}
\newcommand{\Ks}{K_{s}^0}
\newcommand{\pio}{\pi^{0}}
\newcommand{\pip}{\pi^{+}}
\newcommand{\pim}{\pi^{-}}
\newcommand{\ddgamma}{\Dsto(\Do\gamma)\Dob}
\newcommand{\ddpio}{\Dsto(\Do\pio)\Dob}
\newcommand{\mbc}{M_{\mathrm{bc}}}
\newcommand{\pqr}{E_9/E_{25}}


\title{ \quad\\[1.0cm] Evidence of time-dependent $CP$ violation in
 the decay $B^0 \to \Dstp \Dstm$ }


\affiliation{Budker Institute of Nuclear Physics, Novosibirsk}
\affiliation{Chiba University, Chiba}
\affiliation{University of Cincinnati, Cincinnati, Ohio 45221}
\affiliation{The Graduate University for Advanced Studies, Hayama}
\affiliation{Hanyang University, Seoul}
\affiliation{University of Hawaii, Honolulu, Hawaii 96822}
\affiliation{High Energy Accelerator Research Organization (KEK), Tsukuba}
\affiliation{Institute of High Energy Physics, Chinese Academy of Sciences, Beijing}
\affiliation{Institute of High Energy Physics, Vienna}
\affiliation{Institute of High Energy Physics, Protvino}
\affiliation{Institute for Theoretical and Experimental Physics, Moscow}
\affiliation{J. Stefan Institute, Ljubljana}
\affiliation{Kanagawa University, Yokohama}
\affiliation{Korea University, Seoul}
\affiliation{Kyungpook National University, Taegu}
\affiliation{\'Ecole Polytechnique F\'ed\'erale de Lausanne (EPFL), Lausanne}
\affiliation{Faculty of Mathematics and Physics, University of Ljubljana, Ljubljana}
\affiliation{University of Maribor, Maribor}
\affiliation{University of Melbourne, School of Physics, Victoria 3010}
\affiliation{Nagoya University, Nagoya}
\affiliation{Nara Women's University, Nara}
\affiliation{National Central University, Chung-li}
\affiliation{National United University, Miao Li}
\affiliation{Department of Physics, National Taiwan University, Taipei}
\affiliation{H. Niewodniczanski Institute of Nuclear Physics, Krakow}
\affiliation{Nippon Dental University, Niigata}
\affiliation{Niigata University, Niigata}
\affiliation{University of Nova Gorica, Nova Gorica}
\affiliation{Osaka City University, Osaka}
\affiliation{Osaka University, Osaka}
\affiliation{Panjab University, Chandigarh}
\affiliation{Saga University, Saga}
\affiliation{University of Science and Technology of China, Hefei}
\affiliation{Seoul National University, Seoul}
\affiliation{Sungkyunkwan University, Suwon}
\affiliation{University of Sydney, Sydney, New South Wales}
\affiliation{Tohoku Gakuin University, Tagajo}
\affiliation{Tokyo Institute of Technology, Tokyo}
\affiliation{Tokyo Metropolitan University, Tokyo}
\affiliation{Tokyo University of Agriculture and Technology, Tokyo}
\affiliation{IPNAS, Virginia Polytechnic Institute and State University, Blacksburg, Virginia 24061}
\affiliation{Yonsei University, Seoul}

\author{K.~Vervink}\affiliation{\'Ecole Polytechnique F\'ed\'erale
     de Lausanne (EPFL), Lausanne} 
  \author{T.~Aushev}\affiliation{\'Ecole Polytechnique F\'ed\'erale de Lausanne (EPFL), Lausanne}\affiliation{Institute for Theoretical and Experimental Physics, Moscow} 
   \author{O.~Schneider}\affiliation{\'Ecole Polytechnique F\'ed\'erale de Lausanne (EPFL), Lausanne} 
 \author{K.~Arinstein}\affiliation{Budker Institute of Nuclear Physics, Novosibirsk} 
   \author{A.~M.~Bakich}\affiliation{University of Sydney, Sydney, New
   South Wales} 
\author{V.~Balagura}\affiliation{Institute for Theoretical and Experimental Physics, Moscow} 
   \author{E.~Barberio}\affiliation{University of Melbourne, School of Physics, Victoria 3010} 
 \author{A.~Bay}\affiliation{\'Ecole Polytechnique F\'ed\'erale de Lausanne (EPFL), Lausanne} 
   \author{V.~Bhardwaj}\affiliation{Panjab University, Chandigarh} 
   \author{U.~Bitenc}\affiliation{J. Stefan Institute, Ljubljana} 
% ...
\collaboration{The Belle Collaboration}

\begin{abstract}
We report a measurement of the $CP$-odd fraction and the
time-dependent 
$CP$ violation in $B^0 \to \Dstp \Dstm$ decays, using 657 million $B\overline{B}$ events
 collected at the $\Upsilon(4S)$ resonance with the Belle detector at
 the KEKB asymmetric-energy $e^+e^-$ collider.  We measure a $CP$-odd
 fraction of $R_{\perp} = 0.125 \pm 0.043 {\rm (stat)} \pm
 0.023{\rm (syst)}$.  From
 the distributions of the proper-time intervals
 between a $B^0 \to \Dstp \Dstm$ decay and the other $B$ meson in the
 event, we obtain evidence of $CP$ violation with measured parameters $\mathcal{A}_{\Dstp \Dstm} = 0.15
 \pm 0.13 {\rm (stat)} \pm 0.04 {\rm (syst)}$ and $\mathcal{S}_{\Dstp
   \Dstm}= -0.96 \pm 0.25 {\rm (stat)}_{-0.16}^{+0.13} {\rm (syst)}$. 
\end{abstract}


\maketitle

\tighten

{\renewcommand{\thefootnote}{\fnsymbol{footnote}}}
\setcounter{footnote}{0}

In the Standard Model (SM), the irreducible complex phase in the
Cabibbo-Kobayashi-Maskawa (CKM) quark-mixing matrix gives rise to
$CP$-violation~\cite{cabbibo}.  
In an $\Upsilon(4S)$ event, the time-dependent decay rate of a neutral $B$ meson to a $CP$
eigenstate is given by
\begin{eqnarray}  
\mathcal{P}(\Delta t) &=& \frac{e^{-|\Delta t|/\tau_{B^0}}}{4
  \tau_{B^0}} \Big \{1 + q \Big [\mathcal{S} \sin(\Delta m_d \Delta t) +  \mathcal{A} \cos(\Delta m_d \Delta t) \Big ] \Big \}, 
\label{eq:basic}
\end{eqnarray}  
where $q = +1\,(-1)$ when the other $B$
meson in the event decays as a $B^0$ ($\overline{B}^0$), $\Delta t =
t_{CP} - t_{{\rm tag}}$ is the proper-time difference between the two
$B$ decays in the event, $\tau_{B^0}$ is the neutral $B$ lifetime and
$\Delta m_d$ is the mass
difference between the two $B^0$ mass eigenstates.  The $CP$-violating
parameters are defined as
\begin{eqnarray}  
\mathcal{S} = \frac{2 \Im(\lambda)}{|\lambda|^2 + 1}, \quad \mathcal{A}
= \frac{|\lambda|^2 - 1}{|\lambda|^2 + 1}, 
\end{eqnarray} 
where $\lambda$ is a complex observable depending on the $B^0$
  and $\overline{B}^0$ decay amplitudes to the final state and the
  relation between the $B$
  meson mass eigenstates and its flavor eigenstates.
At the quark level the $B^0 \to \Dstp \Dstm$ decay is a $b \to
c\overline{c}d$ transition, where the tree amplitude is CKM-suppressed. The
contribution of penguin diagrams in this decay is estimated to be at
the percent level~\cite{theory}.  If penguin corrections are neglected,
the SM expectations for the $CP$ parameters are $\mathcal{A}_{\Dstp
  \Dstm} = 0$ and $\mathcal{S}_{\Dstp \Dstm} = -\eta_{\Dstp \Dstm} \sin 2
\phi_1$, where $\phi_1 = \arg[-V_{cd}V^*_{cb}]/[V_{td}V^*_{tb}]$ and
$\eta_{\Dstp \Dstm}$ is the $CP$ eigenvalue of  $\Dstp \Dstm$, which
is $+1$ when the decay proceeds through an $S$ or $D$ wave,
or $-1$  for a $P$ wave.  A large measured deviation from this expectation
 can be a sign of new physics~\cite{theory2}.  Recently
Belle reported a $4.1\,\sigma$ $CP$ violation effect in the $B^0 \to D^+
D^-$ decay~\cite{sasa}; $\mathcal{S}$ was found to be consistent with
  $-\sin 2 \phi_1$ whereas the measured $\mathcal{A}$ value indicated $3.2\,\sigma$ direct $CP$
violation, which contradicts the SM and is not
confirmed by BaBar~\cite{babardpldmin}.  This decay contains the
same weak phase transition as $B^0 \to \Dstp \Dstm$, therefore a precise
measurement of the latter is vital for a correct interpretation.
The $CP$-violating parameters as well as the $CP$-odd
fraction in $B^0 \to \Dstp \Dstm$ decays have been measured by both
Belle~\cite{miyake} and BaBar~\cite{babarnew}. Here we report a new
  measurement with more than four times the statistics used in~\cite{miyake}.

This analysis is
based on a data sample containing $657$ million $B \overline{B}$ pairs,
collected with the Belle detector at the KEKB asymmetric-energy  $e^+
e^-$ collider~\cite{kekb} operating at the $\Upsilon(4S)$
resonance. The $\Upsilon(4S)$ meson is produced with a Lorentz
boost $\beta \gamma = 0.425$ nearly along the $z$ axis, defined as the
 direction opposite to that of the positron beam.  Since the $B^0$ and $\overline{B}^0$ are
approximately at rest in the $\Upsilon(4S)$ center-of-mass (CM)
frame, $\Delta t$ can be determined from the displacement in $z$ between
the two decay vertices, $\Delta t \simeq \Delta z / (\beta \gamma c)$,
where $c$ is the speed of light.  

The Belle detector is a large-solid-angle magnetic spectrometer that
consists of a silicon vertex detector (SVD), a 50-layer central drift
chamber (CDC), an array of aerogel threshold Cherenkov counters (ACC),
a barrel-like arrangement of time-of-flight scintillation counters
(TOF), and an electromagnetic calorimeter (ECL) comprised of CsI (Tl)
crystals located inside a superconducting solenoid coil that provides
a $1.5\,{\rm T}$ magnetic field.  An iron flux-return located outside the coil
is instrumented to detect $K_L^0$ mesons and to identify muons (KLM).
A detailed description of the Belle detector can be found
elsewhere~\cite{det}. Two different inner detector configurations were
used.  A first sample of $152 \times 10^6$ $B\overline{B}$
pairs were recorded with a 2.0 cm radius beampipe and a 3-layer silicon vertex detector; for the remaining $505 \times 10^6$ $B\overline{B}$ pairs,
a 1.5 cm radius beampipe, a 4-layer silicon detector, and a
small-cell inner drift chamber were used~\cite{det2}.

Charged particles are reconstructed requiring the
transverse (longitudinal) distance between the track trajectory and the
interaction point (IP) to be less than $2.0\,(4.0)\,{\rm cm}$.  
Neutral pions are
reconstructed from pairs of photons with energies
above $30\,{\rm MeV}$ and with a 
total momentum in the CM system $p_{\gamma \gamma} > 0.1\,{\rm
  GeV}/c$, which are required to have an invariant
mass in the range $119 \ {\rm MeV}/c^2 <M_{\gamma\gamma}<146  \
{\rm MeV}/c^2$.  Neutral kaons are reconstructed via the decay $K^0_s
\to \pip \pim$~\cite{chen}.  The $\pip \pim$ invariant mass is required to be
 within $\pm 9\,{\rm MeV}/c^2$ of the $K^0_s$ mass~\cite{pdg}
 and is constrained in mass and fitted to a common vertex.  
The $\pip \pim$ vertex is required to be displaced from the IP
  in the direction of the pion pair momentum.  The neutral $D$ mesons are reconstructed in the  $ \Km \pip$, $\Km \pip \pio$, $\Km \pip \pip \pim$, $\Ks
\pip \pim$, $\Ks \pip \pim \pio$ and $\Kp \Km$ modes, while $\Dp$ decays are reconstructed in the $\Km \pip \pip$,  $\Ks \pip$, $\Ks \pip \pio$ and $\Kp \Km
\pip$ modes.  Unless specified otherwise charge-conjugated decays are
implied throughout. 

Charged kaons and pions are separated using a likelihood ratio,
$\mathcal{R}_{K/ \pi} =
  \mathcal{L}(K)/(\mathcal{L}(K)+\mathcal{L}(\pi))$, constructed from
ACC information, CDC $dE/dx$ and TOF measurements.   
Charged tracks in 2-prong (3- or 4-prong) vertices are
reconstructed as kaons if $\mathcal{R}_{K /\pi} > 0.1\,
(0.6)$  and as pions when $\mathcal{R}_{K/\pi} < 0.9$.
These requirements have an efficiency of $97\%$ ($85\%$) for kaons in
2-prong (3- or 4-prong) vertices and $98\%$ for pions, 
respectively, with fake rates of $18\%$ ($14\%$) for kaons and $12\%$ for pions.
 The invariant mass of the $D$ candidates must be within
 $\pm 6 \sigma\,
(3 \sigma)$ of the nominal value for 2-prong (3- or 4-prong)
decays, where $\sigma$ is the width of the main component of the channel-dependent $D$ mass resolution obtained from signal Monte Carlo (MC) samples and ranges from
$2.6\,{\rm MeV}/c^2$ to $7.5\,{\rm MeV}/c^2$.  Candidate $\Dstp$
mesons are reconstructed in the $D^0 \pi^+$ and $D^+ \pi^0$ modes. 

The pions from the $D^*$ decays are referred to as slow
pions because of their low momentum.  Slow charged pions are
constrained to originate from the point where the $D$
trajectory intersects the beam profile.  The mass difference,
$\Delta M =| M(\Dst) - M(D)|$ is required to be within 
$ \pm 3\,(2.25)\ {\rm MeV}/c^2$ of the
nominal  value for the $D^0$ ($D^+$) channel. Finally, two oppositely
charged $\Dst$ mesons are combined to form a $B^0$ candidate.
Because of the smaller product branching
fraction and the large background contribution, we do not include $(D^+
\pi^0)(D^- \pi^0)$ combinations. 

The selected $D$ meson
candidates are then
subjected to mass- and vertex-constrained fits to improve their momentum
and vertex resolution.   To discriminate the signal $B$ mesons
  from background, we use the energy difference 
$\Delta E \equiv E_{B}^{{\rm CM}} - E_{{\rm beam}}^{{\rm CM}}$ and the beam-constrained mass $M_{{\rm bc}} \equiv \sqrt{(E_{{\rm beam}}^{{\rm CM}})^2 -
  (p_{B}^{{\rm CM}})^2}$,  where $E_{{\rm beam}}^{{\rm CM}}$ is the beam
energy in the CM system and $E_{B}^{{\rm CM}}$ and
$p_{B}^{{\rm CM}}$ are 
the energy and momentum of the $B$ candidate in the CM system. 
After alle the above selection
requirements are applied, there are on average 1.7 $B^0$ candidates
per event in the large signal region.  This region is defined by  $5.23\,{\rm GeV}/c^2 <
M_{{\rm bc}} < 5.30\,{\rm GeV}/c^2$  and $-0.14\,{\rm GeV} < \Delta E <
0.14\,{\rm GeV}$.  

We choose the $B$ candidate with the
smallest  value of 
\begin{eqnarray}
\chi^2_{{\rm mass}} &=& \sum_{i=1}^2\left(\frac{\Delta M_i-\Delta
 M_i({\rm PDG})}{\sigma_{\Delta M_i}} \right)^2 +
 \sum_{i=1}^2\left( \frac{M(D_i)-M(D_i)({\rm PDG})}{\sigma_{M(D_i)}}
 \right)^2,
\end{eqnarray}
where PDG refers to the world average measurement in~\cite{pdg} and $i$ denotes the two
$D$ mesons.
The $e^+ e^- \to q\overline{q}\,\,(q = u,d,s, {\rm \, and\, }
c)$ background is suppressed by requiring the ratio of the
second- to zeroth-order Fox-Wolfram moments~\cite{r2} to be less than $0.4$.

\begin{figure}
\includegraphics[width= 0.799\textwidth]{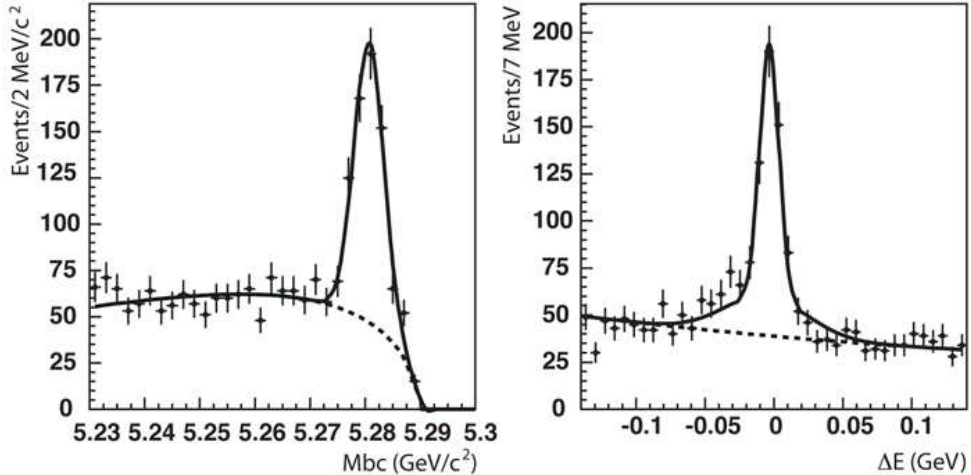}
\caption{(a) The $M_{{\rm bc}}$ distribution for $|\Delta E| < 0.04\,{\rm GeV}$. 
(b) The $\Delta E$ distribution for $M_{{\rm bc}} > 5.27\,{\rm GeV}/c^2$.  The solid curve shows the result of the fit while
  the dotted curve is the background contribution.}
\label{fig:signal}
\end{figure}

We perform an unbinned two-dimensional maximum likelihood fit to the
large signal region in the $M_{{\rm bc}}$ vs. $\Delta E$ plane.  The probability density function (PDF) used to model
the $M_{{\rm bc}}$ distribution is the sum of a signal and background
component.  The signal PDF is described with a Gaussian
 function while the combinatorial background is modeled with an ARGUS
 function~\cite{argus}.  The $\Delta E$ signal distribution is fitted with the
 sum of two Gaussians where the width and mean of the second wide
 Gaussian, as well as the relative fraction of the two Gaussians, are fixed
 to the MC values.  The $\Delta E$ background 
distribution is described with a second-order polynomial.  
Figure~\ref{fig:signal} shows two different projections of the
two-dimensional distribution and fit results.
   We obtain $553 \pm 30$ signal events in the
large signal region.  In the small signal region, defined by $5.27\,{\rm GeV}/c^2 <
M_{{\rm bc}} < 5.30\,{\rm GeV}/c^2$  and $-0.04\,{\rm GeV} < \Delta E <
0.04\,{\rm GeV}$ the signal purity is $55\%$.

\begin{figure}
\includegraphics[width= 0.5\textwidth]{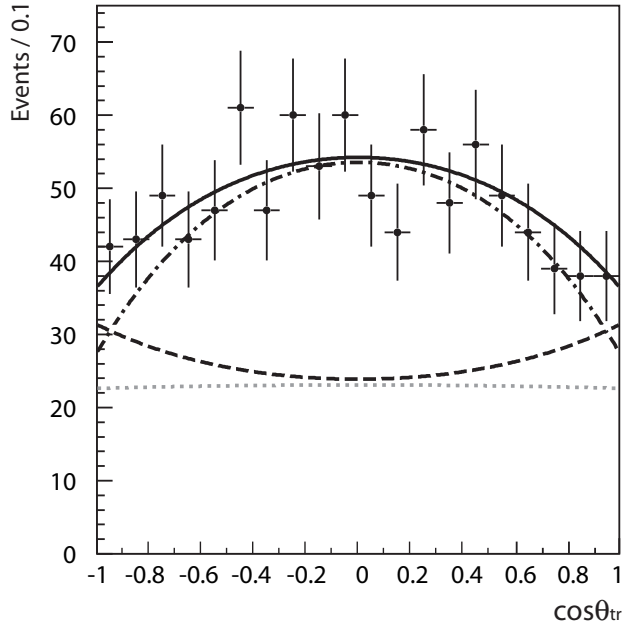}
\caption{The $\cos\theta_{{\rm tr}}$ distribution for events in the
  small signal region,
  the points with error bars represent data.  The solid curve is the result of the fit, the dotted curve shows the background
  contribution. The $CP$-even and $CP$-odd contributions are the dot-dashed and dashed curves, respectively, and are visible above the dotted background curve.}
\label{fig:dstdstpolar_vf}
\end{figure}

To obtain the $CP$-odd fraction we perform a
time-integrated angular analysis in the transversity
basis~\cite{transversity2}.  The differential decay rate as a function
of the transversity angle is
\begin{eqnarray}  
\frac{1}{\Gamma}\frac{d\Gamma}{d\cos\theta_{{\rm tr}}} &=&
   \frac{3}{4} (R_0 + R_{\parallel} )\sin^2\theta_{{\rm tr}} + \frac{3}{2} R_{\perp}\cos^2\theta_{{\rm tr}} 
\label{eq:polthtr}
\end{eqnarray} 
where $R_{0,\parallel}$ and $R_{\perp}$ are the fractions of the
longitudinal, transverse parallel and transverse perpendicular
components in the transversity basis.  $R_{0}$ and $R_{\parallel}$
are the fractions of the $CP$-even polarization while $R_{\perp}$ is
the fraction of the $CP$-odd one.  A one-dimensional fit to 
the $\cos \theta_{{\rm tr}}$ distribution allows the extraction of the $CP$-odd
fraction, where $\theta_{{\rm tr}}$ is the polar angle between the momentum 
of the charged slow pion in its mothers $D^{*}$ rest frame and the normal to the other  $D^{*}$ decay plain.
The measured distribution of $\cos \theta_{{\rm tr}}$ is distorted, in
particular due to the angular resolution of the slow pion.  The shapes of the $CP$-odd and
$CP$-even polarizations are obtained 
from a signal MC sample taking the $R_0/(R_{0} + R_{\parallel})$
fraction from the previous Belle analysis~\cite{miyake}.
The background shape is obtained
from the fit, but limited to be a symmetric polynomial, i.e. $a_{{\rm bkg}} \cdot \cos^2 \theta_{{\rm tr}} + 1$. 
The signal-to-background ratio is determined on an event-by-event basis 
using the $M_{{\rm bc}}-\Delta E$ distribution.  The fit to the large
signal region yields
\begin{eqnarray}  
R_{\perp} &=& 0.125 \pm 0.043 
\label{eq:polthtr2}
\end{eqnarray} 
and $a_{{\rm bkg}} = -0.02 \pm 0.04$.  The fit result is shown in
 Fig.~\ref{fig:dstdstpolar_vf},
superimposed on the $\cos \theta_{{\rm tr}}$ distribution in the small signal region.
This result is compatible with previous Belle and BaBar measurements~\cite{miyake, babarnew}.   

The systematic uncertainty on $R_{\perp}$ is obtained by varying the fixed parameters within
their errors. The
signal efficiency and the $R_0/(R_0 + R_{\parallel})$ parameters give rise to systematic  
uncertainties of $0.003$ and $0.009$, respectively. When varying
  the number of signal events by $\pm 1\sigma$ and 
  the signal shape in $M_{{\rm bc}}$ and $|\Delta E|$ such that the data points in the lower tail in $|\Delta E|$ are well
  described, a systematic uncertainty of $0.003$ is obtained.  A fast MC is 
used to estimate any possible fit bias; we find a small shift of
$0.002$.  Tighter vertex quality cuts lead to a $0.013$ difference in $R_{\perp}$.   Finally, a 
peaking background contribution of $6.6\%$ obtained from the MC is added, to which we conservatively assign a $CP$-odd behavior, leading to a $0.016$ change in the 
central value.  The different contributions are summed in quadrature
to yield a systematic uncertainty of $0.023$ in $R_{\perp}$.  

To determine the $CP$-violating parameters, the signal $B^0$-meson
 decay vertex is reconstructed by fitting the momentum vector
 of the $D$ meson with the beam spot profile. 
No information on the slow pions is used.  After additional
requirements on the number of SVD hits and the vertex fit quality, we obtain $511 \pm 28$ events.


The tag-side decay vertex and the flavor of the tag-side $B$ meson are obtained inclusively from 
properties of particles that are not associated with the
reconstructed $B^0 \to \Dstp \Dstm$ decay~\cite{tag}.  The PDF used to describe
the $\Delta t$ distribution is:
\begin{eqnarray}
\mathcal{P}(\Delta t) &=& \int \Big[ f_{{\rm sig}} \mathcal{P}_{{\rm sig}}(\Delta t') 
+ (1 - f_{{\rm sig}})\mathcal{P}_{{\rm bkg}}(\Delta t') \Big] \cdot R_{{\rm res}}(\Delta t - \Delta t') d(\Delta t'). \\
\end{eqnarray}
The signal fraction, $f_{{\rm sig}}$ and the $CP$-odd probability
are obtained on an 
event-by-event basis, using the previous fits to the $M_{{\rm bc}}$, $\Delta E$ and
$\cos\theta_{{\rm tr}}$ distributions.  The function $\mathcal{P}_{{\rm sig}}$ is given by
Eq.~\ref{eq:basic} and modified event-by-event to incorporate the effect of incorrect flavor
assignment.  A dilution factor,
$[1-2f_{\perp}(\cos\theta_{\rm{tr}})]$ for $\mathcal{S}$ takes into
account the fraction of the $CP$-odd component.  We assume
$\mathcal{S}_{\rm{even}} = - \mathcal{S}_{\rm{odd}}$
($\mathcal{A}_{\rm{even}} = \mathcal{A}_{\rm{odd}}$) and define it as
$\mathcal{S}$ ($\mathcal{A}$).   The tagging quality is parameterized
by a variable $r$ that ranges from $r = 0$ (no flavor discrimination) to
$r = 1$ (unambiguous flavor assignment).  The data is divided into
seven $r$ intervals. The wrong tag fraction
$w_r$, possible tagging performance differences between $B^0$ and $\overline{B}^0$
decays ($\Delta w_r$), and the parameters of the resolution
function $R_{{\rm res}}$ are determined using a high-statistics control
sample of semileptonic and hadronic $b \to c$
decays~\cite{tag, tagrbin}. However, the width of the main Gaussian component of the 
resolution is determined using a $B^0 \to D^{(*)+} D^{(*)-}_{s}$ control sample.  
The parameters of $\mathcal{P}_{{\rm bkg}}(\Delta t)$ are obtained
from a fit to the $\Delta t$ distribution in sideband ($M_{{\rm bc}} <
  5.27\,{\rm GeV}/c^2$) events.   

\begin{figure}
\includegraphics[width= 0.4\textwidth]{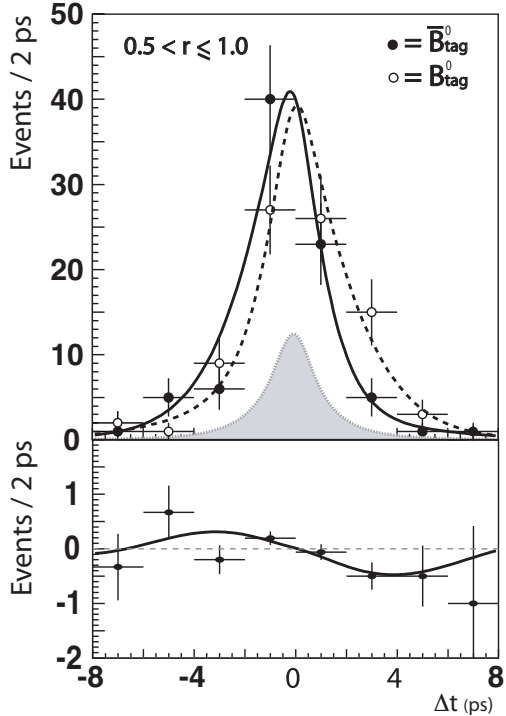}
\caption{Top: $\Delta t$ distribution of well-tagged $B^0 \to \Dstp
  \Dstm$ candidates  ($r > 0.5$) for $q = +1$ and $q = -1$.  The shaded area is the background
  contribution while the solid and dashed curves are the superposition of the
  total PDFs for well-tagged $q=-1$ and $q=+1$ events, respectively.  Bottom: fitted raw asymmetry of the top two distributions.}
\label{fig:cpfit}
\end{figure}

The free parameters in the fit are
$\mathcal{A}_{\Dstp \Dstm}$ and $\mathcal{S}_{\Dstp \Dstm}$; these are determined by
maximizing an unbinned likelihood function for all events in the large
fit region.  The result is:
\begin{eqnarray}
\mathcal{S}_{\Dstp \Dstm} &=& -0.96 \pm 0.25, 
\nonumber \\
\mathcal{A}_{\Dstp \Dstm} &=& +0.15 \pm 0.13,
\end{eqnarray}
with a statistical correlation of $11\%$.  The
significance of $CP$ violation 
using the statistical uncertainty only is $3.4\,\sigma$.  Our measurements of
$\mathcal{S}$ and $\mathcal{A}$ are consistent with the SM expectation for a
tree-dominated $b \to c \overline{c}d$ transition.  The large direct
$CP$ violation measured in $B^0 \to D^+ D^-$~\cite{sasa} is thus not confirmed in
this $b \to c \overline{c}d$ decay mode, in agreement with BaBar's result~\cite{babarnew}.  We define the
raw asymmetry in each $\Delta t$ bin as $(N_{+} - N_{-})/(N_{+} +
N_{-})$, where $N_{+}\, (N_{-})$ is the number of observed candidates
with $q = +1\, (-1)$.  Figure~\ref{fig:cpfit} shows the $\Delta t$ distribution and the raw asymmetry for events with
a good-quality tag ($r > 0.5$) in the smalls signal region.

The systematic uncertainties on the $CP$-violation parameters are
summarized in Table~\ref{table:syserr}. 
\begin{table}[h]
\begin{center}
\caption{Systematic errors on the $CP$-violating parameters
for $B^0 \to \Dstp \Dstm$ decays.}
\begin{tabular}{l  c  c c  }
Source & $\mathcal{A_{\Dstp \Dstm}}$   & \multicolumn{2}{c}{$\mathcal{S}_{\Dstp \Dstm}$}\\
$CP$-odd fraction $R_{\perp}$          &  $\pm 0.004$    &  \multicolumn{2}{c}{$\pm 0.109$} \\
Signal purity and shape        &  $\pm 0.020$    &  \multicolumn{2}{c}{$\pm 0.030$}    \\
Standard resolution function   &  $\pm 0.004$    &  \multicolumn{2}{c}{$^{+0.000}_{-0.102}$} \\
Resolution from control sample    & $\pm 0.002$  &   \multicolumn{2}{c}{$\pm 0.030$} \\
Background shape             &  $\pm 0.000$    &  \multicolumn{2}{c}{$\pm 0.006$}  \\
Fit bias              &  $\pm 0.010$    &  \multicolumn{2}{c}{$\pm 0.031$} \\
$\Delta m_d$, $\tau_{B^0}$               &  $\pm 0.002$    &  \multicolumn{2}{c}{$\pm 0.004$} \\
Flavor tagging               &  $\pm 0.011$    &  \multicolumn{2}{c}{$\pm 0.020$} \\
Vertex cuts               &  $\pm 0.003$    &  \multicolumn{2}{c}{$\pm 0.028$} \\
$\Delta t$ fit range            &  $\pm 0.010$    &  \multicolumn{2}{c}{$\pm 0.004$} \\
Peaking background           &  $\pm 0.010$    & \multicolumn{2}{c}{$^{+0.000}_{-0.027}$} \\
Tag-side interference   &  $\pm 0.034$    &   \multicolumn{2}{c}{$\pm 0.007$} \\
\hline
Total & $\pm$0.044 &   \multicolumn{2}{c}{$^{+0.126}_{-0.164}$} \\
\end{tabular}
\label{table:syserr}
\end{center}
\end{table}
The largest contribution comes from the $R_{\perp}$ fraction, which
only affects $\mathcal{S}_{\Dstp \Dstm}$.
The systematic uncertainty due to the signal-to-background ratio is
 determined by varying the signal yield with $\pm 1 \sigma$, the shape parameters such that the data points in the lower tail in $|\Delta E|$ are well
  described, and the value of $R_{\perp}$ in a correlated way, as the signal purity 
also affects the angular analysis.  $R_{\perp}$ is varied by $0.003$, which is the
systematic error in $R_{\perp}$ due to the signal purity and shape.  
The contribution of the resolution function and
the background shape to the systematic error is estimated by varying each
parameter by $\pm 1 \sigma$.    Varying the
resolution parameters moves $\mathcal{S}_{\Dstp \Dstm}$ further away
from zero.   A fast MC is used to estimate the bias of the $CP$ violating
parameters for the measured values. 
The $\Delta m_d$ and
$\tau_{B0}$ parameters are varied around their world
averages~\cite{pdg}.  
Systematic errors due to uncertainties in
wrong tag fractions are estimated by varying the parameters $w_l$ and
$\Delta w_l$ in each $r$ region by their $\pm 1 \sigma$ errors.
The vertex quality cut is changed to
$\xi < 125$ and the effect is included in the table.  
The $\Delta t$ fit range is changed from $\Delta t <
70\,{\rm ps}$ to $\Delta t <10\,{\rm ps}$. A peaking background
contribution is added with no $CP$ violation. Finally, the
tag-side interference uncertainty is included~\cite{phys}.  The different 
sources are added in quadrature to yield $\pm 0.04$ for
$\mathcal{A}$ and $_{-0.16}^{ +0.13}$ for $\mathcal{S}$, reducing the
significance of $CP$ violation to $3.1\,\sigma$.

We performed various cross-checks such as a fit to the
$CP$ asymmetries of the control sample $B^0 \to D^{(*)+}
D^{(*)-}_{s}$, which gives $A = -0.02 \pm 0.03
{\rm (stat)}$ and $S = -0.07 \pm 0.04 {\rm (stat)}$; these values are consistent with no $CP$
asymmetry.  The lifetime fit to the $B^0 \to \Dstp \Dstm$ sample is consistent with the world average
value~\cite{pdg}.  

In summary, we have performed new measurements of the $CP$-odd
fraction $R_{\perp} = 0.125 \pm 0.043 {\rm (stat)} \pm
 0.023{\rm (syst)}$ and $CP$-violation parameters $\mathcal{A}_{\Dstp \Dstm} = \,\,\,0.15 \pm 0.13 {\rm (stat)} \pm 0.04 {\rm (syst)}$ and
$\mathcal{S}_{\Dstp \Dstm} = -0.96 \pm 0.25 {\rm (stat)}_{- 0.16}^{+0.13}
 {\rm (syst)}$ for the decay $B^0 \to \Dstp \Dstm$ using $657 \times
 10^6\, B\overline{B}$ events.    We obtain evidence of
 $CP$ violation with $3.1\,\sigma$ significance including systematic uncertainties.  These measurements
 are consistent with and supersede our previous results~\cite{miyake}.
 They are also in agreement with the SM prediction for $b\to c$ tree
 amplitudes and do not confirm the large direct $CP$ violation seen in the
 $B^0 \to D^+ D^-$ decay.

We thank the KEKB group for the excellent operation of the
accelerator, the KEK cryogenics group for the efficient
operation of the solenoid, and the KEK computer group and
the National Institute of Informatics for valuable computing
and SINET3 network support.  We acknowledge support from
the Ministry of Education, Culture, Sports, Science, and
Technology (MEXT) of Japan, the Japan Society for the 
Promotion of Science (JSPS), and the Tau-Lepton Physics 
Research Center of Nagoya University; 
the Australian Research Council and the Australian 
Department of Industry, Innovation, Science and Research;
the National Natural Science Foundation of China under
contract No.~10575109, 10775142, 10875115 and 10825524; 
the Department of Science and Technology of India; 
the BK21 and WCU program of the Ministry Education Science and
Technology, the CHEP SRC program and Basic Research program (grant No.
R01-2008-000-10477-0) of the Korea Science and Engineering Foundation,
Korea Research Foundation (KRF-2008-313-C00177),
and the Korea Institute of Science and Technology Information;
the Polish Ministry of Science and Higher Education;
the Ministry of Education and Science of the Russian
Federation and the Russian Federal Agency for Atomic Energy;
the Slovenian Research Agency;  the Swiss
National Science Foundation; the National Science Council
and the Ministry of Education of Taiwan; and the U.S.\
Department of Energy.
This work is supported by a Grant-in-Aid from MEXT for 
Science Research in a Priority Area ("New Development of 
Flavor Physics"), and from JSPS for Creative Scientific 
Research ("Evolution of Tau-lepton Physics").


\begin{thebibliography}{99}
\bibitem{cabbibo} M.~Kobayashi and T.~Maskawa, Prog. Theor. Phys. {\bf 9}, 652 (1973).
\bibitem{theory} X.Y.~Pham {\em et al.}, Phys. Lett. B {\bf 458}, 375
  (1999).
\bibitem{theory2} Y.~Grossman {\em et al.}, Phys. Lett. B {\bf 395}, 241
  (1997).
\bibitem{sasa} S.~Fratina {\em et al.} (Belle Collaboration),
  Phys. Rev. Lett. {\bf 98}, 221802 (2007).
\bibitem{babardpldmin} B.~Aubert et al. (BaBar Collaboration),
  Phys. Rev. Lett. {\bf 99}, 071801 (2007).
\bibitem{miyake} H.~Miyake {\em et al.} (Belle Collaboration),
  Phys. Lett. B {\bf 618}, 34 (2005).
\bibitem{babarnew} B.~Aubert {\em et al.} (BaBar Collaboration),
  Phys. Rev. D {\bf 79}, 032002 (2009).
\bibitem{kekb} S.~Kurokawa and E.~Kikutani, Nucl. Instrum. Methods
  Phys. Res., Sect. A {\bf 499}, 1 (2003), and other papers included
  in this volume.
\bibitem{det} A.~Abashian {\em et al.} (Belle Collaboration), Nucl. Instrum. Methods
  Phys. Res., Sect. A {\bf 479}, 117 (2002).
\bibitem{det2} Z.~Natkaniec (Belle SVD2 Group), Nucl. Instrum. Methods
  Phys. Res., Sect. A {\bf 560}, 1 (2006).
\bibitem{chen} K-F.~Chen et al. (Belle Collaboration), Phys. Rev. D {\bf 72}, 012004 (2005).
\bibitem{pdg} C.~Amsler {\em et al.} (Particle Data Group), Phys. Lett.
  B {\bf 667}, 1 (2008). 
\bibitem{r2} G.C.~Fox, S.~Wolfram,  Phys. Rev. Lett. {\bf 41},
  1581 (1978).
\bibitem{argus} H.~Albrecht {\em et al.} (ARGUS Collaboration),  Phys. Rev. Lett. B {\bf 241},
  278 (1990).
\bibitem{transversity2} The BaBar Collaboration Physics Book, edited
  by P.~F.~Harrison and H.~R.~Quinn, SLAC-R504, pp. 213--220 (1998).
\bibitem{svdhits} H.~Tajima {\em et al.},  Nucl. Instrum. Methods
  Phys. Res., Sect. A {\bf 533}, 370 (2004).
\bibitem{tag} H.~Kakuno {\em et al.},  Nucl. Instrum. Methods
  Phys. Res., Sect. A {\bf 533}, 516 (2004).
\bibitem{tagrbin} K.~Abe {\em et al.} (Belle
  Collaboration), Phys. Rev. D {\bf 71}, 072003 (2005).
\bibitem{phys} O.~Long, M.~Baak, R.~N.~Cahn and D.~Kirkby,
  Phys. Rev. D {\bf 68}, 034010 (2003).
\end{thebibliography}
\end{document}